\documentclass[letterpaper]{article} 
\usepackage[draft]{aaai25}  
\usepackage{times}  
\usepackage{helvet}  
\usepackage{courier}  
\usepackage[hyphens]{url}  
\usepackage{graphicx} 
\usepackage{enumitem}
\usepackage{natbib}  
\usepackage{caption} 
\usepackage[linesnumbered,ruled,vlined]{algorithm2e}

\usepackage{color, soul, booktabs, subcaption, makecell, multirow, tabularx, dsfont}
\usepackage[short, nocomma]{optidef}
\allowdisplaybreaks
\DeclareMathOperator*{\expect}{\mathbf{E}}

\title{Contextual Stochastic Optimization for School Desegregation Policymaking}

\author {
    Hongzhao Guan\textsuperscript{\rm 1},
    Nabeel Gillani\textsuperscript{\rm 2, 3},
    Tyler Simko\textsuperscript{\rm 4},
    Jasmine Mangat\textsuperscript{\rm 2},
    Pascal Van Hentenryck\textsuperscript{\rm 1}
}
\affiliations {
    \textsuperscript{\rm 1}H. Milton Stewart School of Industrial and Systems Engineering, Georgia Institute of Technology\\
    \textsuperscript{\rm 2}College of Arts, Media and Design, Northeastern University\\
    \textsuperscript{\rm 3}D'Amore-McKim School of Business, Northeastern University\\
    \textsuperscript{\rm 4}Department of Government, Harvard University\\
    hguan7@gatech.edu,
    n.gillani@northeastern.edu, tsimko@g.harvard.edu,
    j.mangat@northeastern.edu,
    pascal.vanhentenryck@isye.gatech.edu
}

\begin{document}

\maketitle

\begin{abstract}

Most US school districts draw geographic ``attendance zones'' to
assign children to schools based on their home address, a process that
can replicate existing neighborhood racial/ethnic and socioeconomic
status (SES) segregation in schools. Redrawing boundaries can reduce
segregation, but estimating expected rezoning impacts is often challenging because
families can opt-out of their assigned schools. This paper seeks to 
alleviate this societal problem by developing a joint
redistricting and choice modeling framework, called {\em redistricting
with choices (RWC)}. The RWC framework is applied to a large US
public school district to estimate how redrawing elementary school
boundaries might realistically impact levels of
socioeconomic segregation. The main methodological contribution of
RWC is a contextual stochastic optimization model that aims to minimize
district-wide segregation by integrating rezoning constraints
with a machine learning-based school choice model. The study finds that RWC
yields boundary changes that might reduce segregation by a substantial
amount (23\%) -- but doing so might require the re-assignment of a
large number of students, likely to mitigate re-segregation that
choice patterns could exacerbate. The results also reveal that
predicting school choice is a challenging machine learning problem.
Overall, this study offers a novel practical framework that both
academics and policymakers might use to foster more diverse and
integrated schools.

\end{abstract}

\begin{links}
    \link{Code}{https://github.com/Plural-Connections/AAAI-2025-Redistricting-with-Choices}
\end{links}

\section{Introduction}
\label{sect:introduction}

Schools in the United States remain highly segregated by race and
Socio-Economic Status (SES), despite evidence that integration reduces
achievement gaps \cite{reardon2018testgaps, reardon2019separate,
  billings2013cms, johnson2019, johnson2011desegregation,
  wells1994integration}. School segregation is largely driven by high
levels of residential segregation across race and SES, as the vast
majority of public school students attend the school they are
geographically assigned to attend via the ``attendance zones'' set by
their local school district~\cite{monarrez2021}.

The practical difficulty of redrawing attendance boundaries has
inspired academic research on automated redrawing methods targeting a
number of goals, including
integration~\cite[e.g.][]{gillani2023redrawing}. However, changing
student assignment policies to foster diverse schools often induces
changes in how families subsequently pick schools for their
children---sometimes yielding ``White flight'', or decisions among
White and affluent families to opt-out of their newly-assigned
schools~\cite{reber2005flight,nielsen2020denmark,macartney2018boards}. {\em
  Better anticipating such responses prior to changing boundaries poses a
  fundamental challenge to designing policies that achieve integration
  goals in practice.} In other words, the redrawing task requires an
interplay between two sub-tasks: choices are first determined
conditional on a possible boundary change (``redistricting''), and
then a redistricting is produced conditional on those
choices. Researchers have previously developed redistricting
algorithms for updating school boundaries and subsequently estimated
how families might select schools in response to changes, but this
model did not factor in school selections to inform the boundary
changes themselves~\cite{allman2022}.

This paper aims to address this challenge: it develops a novel
joint redistricting and choice modeling framework, called {\em
  redistricting with choices (RWC)}. The RWC framework proposes {\em a
  contextual stochastic constraint program} that minimizes
district-wide dissimilarity, and integrates the rezoning constraints
and a school choice model. RWC derives the school choice model using a
machine learning model, leveraging features of students, Census Blocks, and
the district schools.

RWC was evaluated on real data from a large public US school district
--- the Winston-Salem/Forsyth County Schools in North Carolina, which
serves over 50,000 students (22,000 of which are enrolled in
elementary schools). The study reveals that the RWC yields an expected
23\% decrease in segregation while rezoning more Census Blocks and
students than baseline choice models. This additional rezoning is
likely necessary to offset a predicted tendency for more students to
opt-out of their zoned schools (and perhaps increase segregation as a
result). The study also shows that predicting school choice is a
challenging multi-class classification problem (F1 score:
0.73). Fortunately, 87\% of the time, the classifier identifies a
student's ground-truth school among its top three most probable
predictions.

The main contributions of the paper can be summarized as follows.
\begin{enumerate}[leftmargin=*, parsep=0pt, itemsep=0pt, topsep=4pt]
    \item To the authors' knowledge, RWC is the first joint school
      redistricting optimization model where school choice
      predictions inform integration-promoting boundary changes.
    \item RWC leverages contextual stochastic optimization, and its
      approximation through the Sample Average Approximation method,
      to obtain a practical tool to inform the design of school
      boundaries that foster integration.
    \item Experimental results on a large school district show that
      RWC can substantially reduce segregation, but typically
      re-assigns more students than other methods.
\end{enumerate}

\noindent
Together, these contributions demonstrate the important role AI can
play in supporting decision-makers seeking to advance equitable access
to quality education.

\section{Related Work} \label{sect:literature}

Much prior work from AI and other fields has demonstrated how attendance zones might be redrawn to reduce racial segregation while respecting local constraints like minimizing travel time to schools \cite[e.g.][]{smilowitz2020, heckman1969, liggett1973, clarke1968, holloway1975interactive, diamond1987, gillani2023redrawing}. Such tools have also been adopted by academics working in partnership with districts~\cite[e.g.][]{allman2022}. Still, implementing them in practice requires careful thought and attention to the goals of policymakers and community members. In general, redistricting is a socio-politically fraught topic, particularly when issues of diversity are at stake, because parents fear that boundary changes might impact friend groups~\cite{bridges2016eden, baltimore2019}, home valuations~\cite{kane2005housing,bridges2016eden, black1999housing}, school quality~\cite{zhang2008flight}, and many other factors. Beyond this paper, the researcher-practitioner partnership that this work is a part of aims to attend to these concerns while also fostering integration.

Researchers have also previously sought to predict school choice. Anticipating school choice is important for planning teams in school districts. Prior work on predicting school choice has developed (mostly linear) demand choice models in centralized school assignment settings~\cite[.e.g][]{pathak2017demand,shi2022optimal}. However, these studies have not applied such models in the context of simultaneously adjusting geographic assignment policies (like attendance boundaries). A notable exception is~\cite{allman2022}, yet as noted earlier, this study is still preliminary given it estimates school choices post-redistricting instead of factoring them into boundary changes. This paper aims at advancing this work by estimating family choices and using those estimates to shape the boundary change process.

Note that researchers in other settings have explored the
integration of choice models in optimization problems
(e.g., optimizing transportation network and routes, which requires
designing transit network conditional on riders' choices, and vice
versa~\cite{basciftci2022transport,guan2024transport}). Such
models often computationally challenging in practice.

\section{Research Setting}

\begin{figure}[!ht]
\centering
\includegraphics[width=\linewidth]{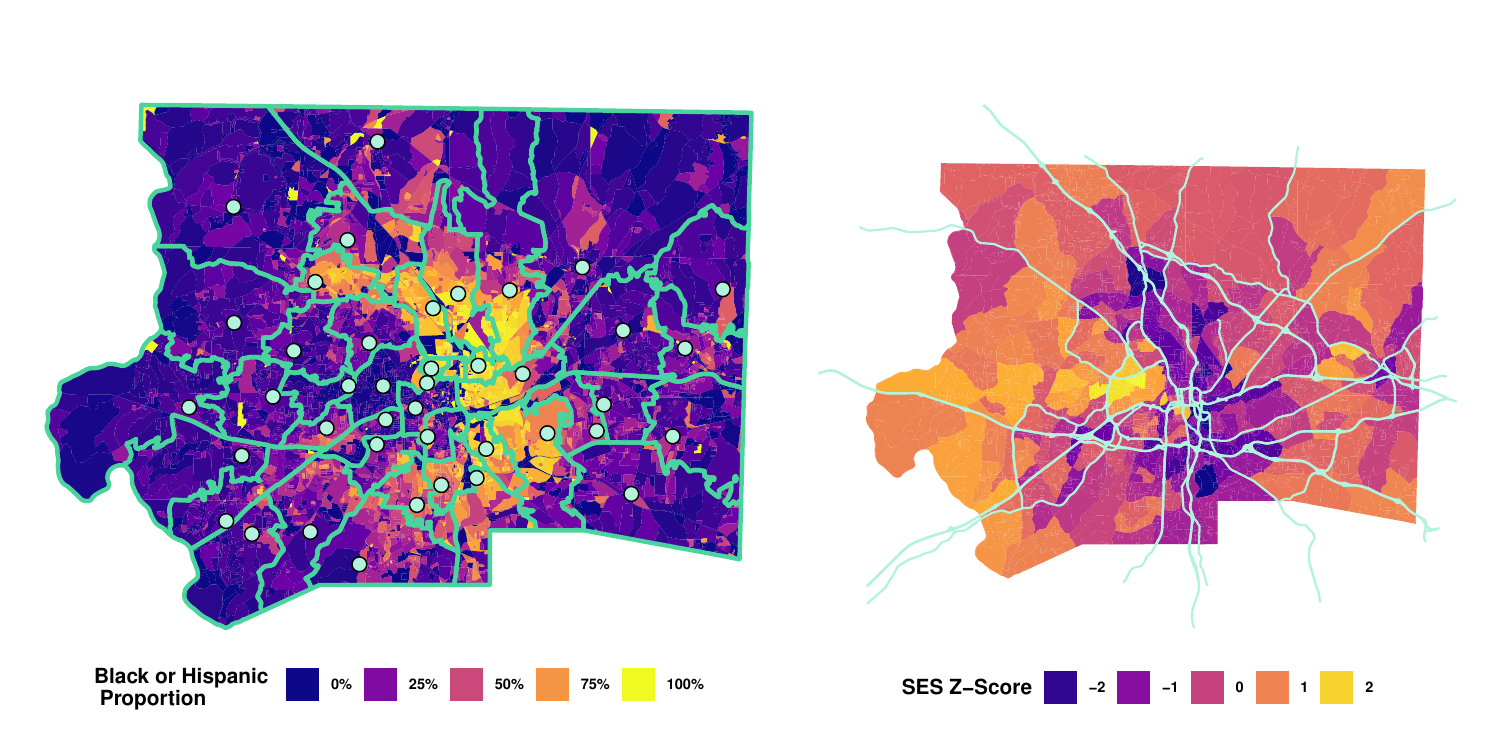}
\caption{Maps of WS/FCS. Left shows race/ethnicity (computed via~\cite{censable}); right shows the distribution of the continuous, geographically-defined Socio-Economic Status (SES) index measure values (described more in Section~\ref{sect:problem_method}).}
\label{fig:ws_fcs_map}
\end{figure} 

The research partner for this study is Winston-Salem/Forsyth County Schools (WS/FCS), a public school district serving students from pre-kindergarten to twelfth grade in North Carolina. With more than 50,000 students, WS/FCS is the 4th largest school district in North Carolina and the 81st largest in the United States. WS/FCS students are racially diverse, with approximately 33\% White students, 30\% Black students, 30\% Hispanic students, 6\% reporting multiple races, and 3\% Asian students. The district operates 81 schools (42 elementary). Despite district-wide diversity, residential and school enrollment patterns are still highly segregated. Figure \ref{fig:ws_fcs_map} visualizes this segregation, highlighting patterns that make WS/FCS one of the most segregated districts in the state. As a recent recipient of a US Department of Education's ``Fostering Diverse Schools'' grant~\cite{jacobson2023fostering}, there is renewed interest in WS/FCS to explore how boundary changes might enhance SES diversity in schools. These would be the first significant changes to district boundaries since the end of court-ordered desegregation in the 1990s~\cite{wsfoundation}. This paper names the district collaborator (with their permission) to ground the empirical explorations and findings in relevant historical and present-day context and inform more thoughtful applications of AI for social impact. 

Most WS/FCS students attend their zoned school, and therefore, the focus of the Fostering Diverse Schools work is largely on redrawing boundaries based on underlying patterns of residential segregation, without trying to anticipate school choice. Still, the district has an active choice program, meaning exploring how choice patterns might alter the effects of boundary changes is essential for the sound implementation of new policies. WS/FCS families are allowed to choose either the school they are residentially (default) assigned to; a magnet school; any school inside a pre-set, contiguous group of schools called a "choice zone" (with guaranteed transportation); or any school outside of their choice zone (in which case, they must provide their own transportation). The district does not use deferred acceptance to centralize student-school matching. Typically, students are able to select any school they wish to attend (and in cases of schools being oversubscribed, the district is usually still able to accommodate them). Therefore, this paper does not simulate deferred acceptance, top-trading cycles, or another typical centralized student assignment mechanism; students are simply assigned to schools by sampling their most likely choice from a probability distribution over schools. 

\section{Problem and Methodology}
\label{sect:problem_method}

\begin{figure*}
\centering
\includegraphics[width=.78\linewidth]{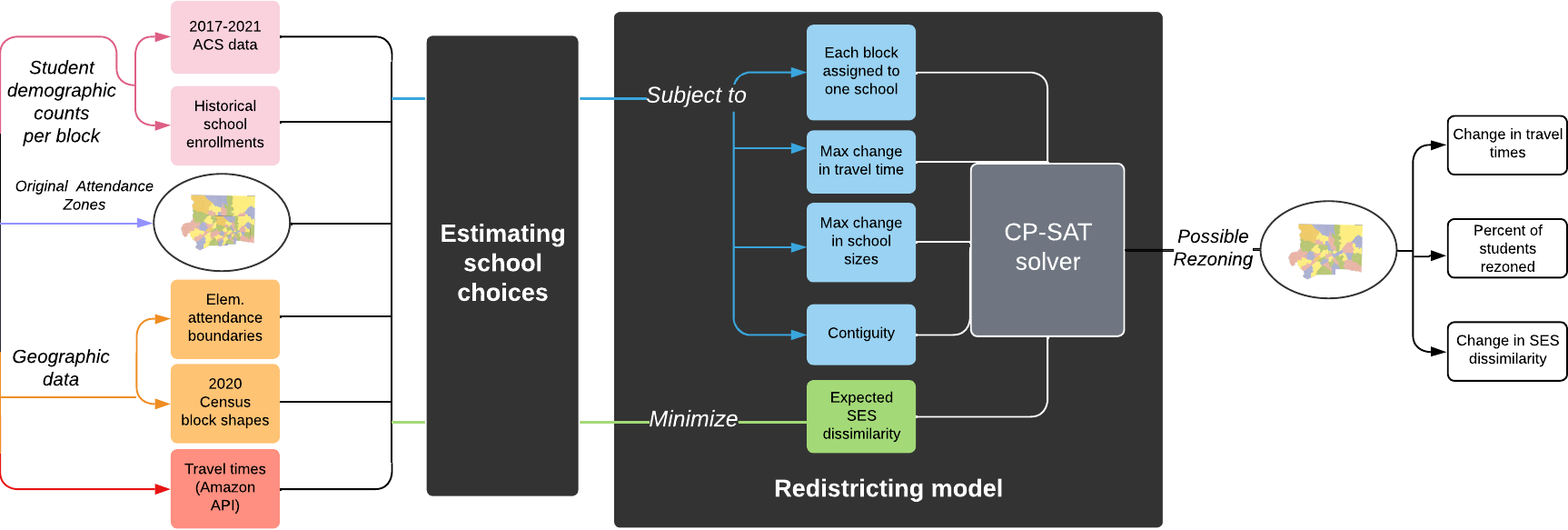}
\caption{Diagram for the RWC framework, adapted from~\cite{gillani2023redrawing}.}
\label{fig:model_overview}
\end{figure*}

This section presents the problem setting and the methodological
approach of the paper. The overall goal is to choose a zoning for a
public school district that minimizes school dissimilarities with
respect to a group of students. The main methodology contribution is
to include a choice model into the optimization, i.e., capturing how
students will respond to changes in their allocated schools. From a
technical standpoint, the methodology uses contextual stochastic
optimization \cite{sadana2024survey} as a framework and constraint
programming for modeling and solving the resulting optimization
problem. Figure~\ref{fig:model_overview} summarizes the overall
methodology, and Table~\ref{tab:nomenclature} summarizes the notation
used in the paper.

\begin{table}[!t]
\centering
\tiny
\begin{tabularx}{\columnwidth}{ l X}
\toprule
\textbf{Notation} & \textbf{Definition} \\
\midrule
\textbf{Sets:} & \\
$\mathcal{SH}, \mathcal{SH}_{mgt}$ & Schools and magnet schools considered in this study ($\mathcal{SH}_{mgt} \subseteq \mathcal{SH}, |\mathcal{SH}| = 41, |\mathcal{SH}_{mgt}| = 8$). \\
$\mathcal{SH}_r$ & The nearest $r$ schools to a student's residential block.\\$\mathcal{B}$ & Census Blocks considered in this study ($|\mathcal{B}| = 6373$). \\
$\mathcal{Z}$ & The set of all feasible assignments. \\
\textbf{Vectors:} & \\
$\mathbf{f_n}, \mathbf{d_n}$ & The static and dynamic feature vectors of student $n$. \\
$\mathbf{x_n}$ & The contextual information and feature vector of student $n$, concatenated by $\mathbf{f_n}$ and $\mathbf{d_n}$. \\
$\mathbf{y}$ & The ground-truth labels, indicating the schools students actually attend in the dataset, i.e., $\mathbf{y}_n = \bar{s}_n^{actual}$. \\
$\mathbf{z}$ & A zoning, constructed by decision variables $z_{b,s}$. \\
$\mathbf{S}$ & A vector of random variables $S_n$. \\
\textbf{Decision Variables:} & \\
$z_{b,s}$ & Binary variable indicates if a census block $b$ is zoned to school $s$. \\
$d$, $d^{i}$ & Continuous variable indicates the district-wide dissimilarity, dynamically based on $\mathbf{z}$. $i$ indicates scenarios for SAA.\\ 
$c_{s}$, $c_{s}^i$ & Integer variable indicates the number of students in school $s$, dynamically based on $\mathbf{z}$. $i$ indicates scenarios for SAA. \\
$g_{s}$, $g_{s}^i$ & Integer variable indicates the number of lower-SES students in school $s$, dynamically based on $\mathbf{z}$. $i$ indicates scenarios for SAA. \\
\textbf{Parameters:} & \\
$I$ & The number of SAA (Sample Average Approximation) scenarios.\\
$\tau$ & The maximum increment ratio on travel time, $\tau \in [0, 1]$. \\
$\alpha$ & The maximum change ratio in school population, $\alpha \in [0, 1]$. \\
\textbf{Constants:} & \\
$N$ & The number of student in the study ($N = 22,302$). \\
$\bar{g}_{total}$ & The number of student in the target group. In this paper, the target group is students in lower-SES category ($\bar{g}_{total} = 9,039$). \\
$\bar{e}_n$ & The SES category of student $n$, determined by their residential area. $\bar{e}_n = 0$ means student $n$ belongs to the lower-SES category. \\
$\bar{t}_{b, s}$ & The estimated driving time from census block $b$ to school $s$. \\
$\bar{b}_n$ & The residential census block of student $n$. \\
$\bar{c}_s$ & The number of current students in school $s$, based on realistic data. \\
$\bar{s}_{b}$ & The current zoned school to  block $b$, based on realistic data. \\
$\bar{s}_n^{zone}$ $\bar{s}_n^{actual}$ & The ground-truth zoned (residentially-assigned school) and actual schools for student $n$, respectively, based on realistic data.\\ 
\textbf{Others:} & \\
$\mathit{S}_n$ & Random variable indicating the choice of school of student $n$. \\
$s_n^{zone}$ & A variable that is used to indicate the zoned school for student $n$, useful for constructing $\mathbf{x}_n$.\\ 
$A_n^i(\mathbf{z})$ &  The school chosen by student $n$ for zoning $\mathbf{z}$. $i$ indicates scenarios for SAA. \\
$\mathcal{C}$, $\mathcal{C}^{f}$, $\mathcal{C}^{rb}$, $\mathcal{C}^{ml}$ & Choice models employed in this study, ``f'', ``rb'', and ``ml'', stand for follow, rule-based, and machine learning, respectively \\ 
$\mathcal{C}^{ml\text{-}l}$, $\mathcal{C}^{ml\text{-}x}$ & Continued from the above, machine learning based choice models, ``l'' and ``x'' stand for multinomial-logit and XGboost, respectively \\ 
\bottomrule
\end{tabularx}
\caption{Notation used throughout this study.}
\label{tab:nomenclature}
\end{table}

\subsection{Data and Notations}

The redistricting problem seeks to assign Census Blocks to schools to
minimize some metrics. Such an assignment is called a {\em zoning}. A
census block is the smallest geographic area determined and publicly
released by the United States Census Bureau. Each census block
typically contains many households, but it can have no population and
only include geographical features such as rivers and mountains.

Four years (2019---2023) of anonymized, geocoded student data were obtained from Winston-Salem/Forsyth
County Schools through
a signed data-sharing agreement, with University IRB approval. During
the 2022-2023 school year, the district operated 41 public elementary
schools across 6,373 Census Blocks, and enrolled
22,302 students.

The set of Census Blocks is denoted by $\mathcal{B}$ and the set of
schools by $\mathcal{SH}$. In the following, a bar notation, e.g.,
$\bar{t}_{b,s}$, represents the value of a quantity in the dataset.
$\bar{s}_b$ denotes the school assigned of
each census block $b$ in the dataset zoning. The census block of
student $n$ is denoted by $\bar{b}_n$. 
The school assigned to student $n$, denoted by $\bar{s}_n^{zone}$, can
be derived from $\bar{b}_n$ and $\bar{s}_b$, However, students are allowed
to opt-out to any of 40 other public schools in the district, and
$\bar{s}_n^{actual}$ represents the school in which the student is
actually enrolled and $\bar{c}_s$ represents the actual student
enrollment at each school $s$. Each student $n$ is assigned a
Socio-Economic Status (SES) index, denoted by $\bar{e}_n$, based on their
residential area. This index is categorized into three levels: lower
(0), medium (1), and higher (2). Its definition is presented in more detail in~\cite{gillani2024raceses} and is similar to one used by peer
districts~\cite{quick2016cps, hawkins2018sanantonio}. At a high level, the measure is created by first producing a geographically-defined, continuous index measure comprised of a number of American Community Survey variables (median household income, adult educational attainment levels, etc.), and then thresholding this continuous measure to create the discrete categories described above: students living in Census Blocks in the bottom, middle, and top third of the distribution for this measure are considered lower, medium, or higher-SES, respectively.

\subsection{Decision Variables and Objective Function}

The decision variables capture the proposed zoning, i.e., binary
variable $z_{b,s}$ represents whether block $b$ is assigned to school
$s$ are represented by binary decision variables. This collection of
decision variables is collected in a zoning vector $\mathbf{z}$. The
auxiliary variables $c_s$ and $g_s$ represent the number of students
and the number of students with lower-SES scores assigned to school
$s$. Zonings must satisfy several important constraints that are
described later in the paper, and the set of feasible zonings is
denoted by $\mathcal{Z}$.

The optimization model aims at finding a zoning that minimizes the
district-wide dissimilarity metric \cite{massey1988segregation}, i.e.,
\begin{small}
\begin{equation}
\label{eq:ses_dis}
        d = \frac{1}{2}~\sum\limits_{s \in S} ~\bigg| \frac{g_s}{\bar{g}_{total}} - \frac{c_s - g_s}{N - \bar{g}_{total}} \bigg|
\end{equation}
\end{small}
where constant $\bar{g}_{total}$ represents the total number of
students with lower-SES scores across all schools (and so, the objective is to minimize the segregation of lower-SES students from their medium and higher-SES counterparts). The dissimilarity
index of a school $s$ ranges from 0 (perfect integration---the
demographic mix of each school reflects district-wide proportions) to
1 (total segregation---all lower-SES students in the ditrict
attend a single school). 

\subsection{Contextual Stochastic Optimization (CSO)}

One of the key challenges of the redistricting problem is to capture
student choices for schools. For instance, a student might attend
school 1 when assigned school 1, but opt-out to school 3 when assigned
to school 2. These decisions, of course, are not available directly; to
overcome this limitation, this paper explores the use of machine
learning and contextual stochastic optimization.

Let $S_n$ be a random variable representing the school student $n$
attends for a proposed zoning and $\mathbf{S}$ be the vector of these
random variables. For a zoning $\mathbf{z}$, a realization of
$\mathbf{S}$ defines the auxiliary variables $c_s$ and $g_s$ defined
earlier and hence the district-wide dissimilarity metric. This is
captured by the function $\textit{dis-score}(\mathbf{s})$ which
computes the district-wide dissimilarity score from a realization of
$\mathbf{S}$.

The redistricting optimization problem can then be formalized as
the following stochastic optimization problem: 
\begin{equation}
\label{eq:cso}
         \min\limits_{\mathbf{z} \in \mathcal{Z}}
         \expect\limits_{\Pr[\mathbf{S} | \mathbf{z}]} [\textit{dis-score}(\mathbf{S})]
\end{equation}
It assumes that students independently choose schools:
\begin{equation}
\label{eq:independent}
        \Pr[ \mathbf{S} \mid \mathbf{z} ] = \prod\limits_{n = 1, ..., N} \Pr[ S_n  \mid \mathbf{z}]
\end{equation}
Again, the distribution $\Pr[ S_n | \mathbf{z}]$ is not readily available. In reality, students may influence one another, so future work should explore lifting this independence assumption.

To approximate \eqref{eq:cso}, CSO combines contextual information about students and schools. In particular, the contextual information
of student $n$, denoted by $\mathbf{x}_n$, comprises personal
information, data from their residential census block, as well as the
current school-related information (e.g., composition of the student
body). The distribution $\Pr[ S_n | \mathbf{z}]$ is then approximated
by $\Pr[ S_n | \mathbf{z}, \mathbf{x}_n]$ and the CSO problem becomes:
\begin{equation}
\label{eq:cso_independent}
         \min\limits_{\mathbf{z} \in \mathcal{Z}}
         \expect\limits_{\prod\limits_{n} \Pr[ S_n  \mid \mathbf{z}, \mathbf{x}_n]} [\textit{dis-score}(\mathbf{S})]
\end{equation}
\noindent
Note that, in this formulation, school choices are solely based on
information available before actual school attendance. In other words,
$\mathbf{x}_n$ is only created using historical data from the 2019-2020 through 2021-2022 school years and the status-quo zoning information for the
2022-2023 school year ($\bar{s}_n^{zone}$). Those features capture
information that previous studies, such as ~\cite{pathak2017demand},
have shown to provide valuable insights into students' and families'
school choices. This includes details such as the programs offered by
schools, school ratings, travel times to schools, and other relevant
factors.

The CSO problem \eqref{eq:cso} is approximated with the Sample Average
Approximation (SAA) method~\cite{kleywegt2002sample}. The solution
method generates $I$ scenarios, each of which defines a choice
function for each student. More precisely, for each student $n$ and
each scenario $i$, $A^i_n(\mathbf{z})$ denotes the school chosen by
student $n$ for zoning $\mathbf{z}$. The optimization becomes:
\begin{equation}
\label{eq:saacso}
         \min\limits_{\mathbf{z} \in \mathcal{Z}}
         \sum\limits_{i \in [I]} [\textit{dis-score}((A^i_1(\mathbf{z}),\ldots,A^i_N(\mathbf{z}))]
\end{equation}
Since each student chooses their school independently, this paper also
uses $A^i_n(s)$ to denote the school chosen by student $n$ when
assigned to school $s$.

\subsection{The Constraint Programming Model} \label{subsect:redistricting}

Figure \ref{fig:cp} depicts a constraint programming model for
formulation \eqref{eq:saacso}, where $d^i$, $c_s^i$,
and $g_s^i$ represent the auxiliary variables computed for a scenario
$i$ given a proposed zoning. The model features the following
constraints:

\begin{enumerate}[leftmargin=*, parsep=0pt, itemsep=0pt, topsep=4pt]

\item \textbf{School Sizes and Population Increases}.
  Equations~\eqref{eq_cp:school_population_dynamic}
  and~\eqref{eq_cp:school_ses_population_dynamic} indicate how the
  total ($c_s^i$) and lower-SES populations ($g_s^i$) of schools are
  derived based on the value of
  $z_{b,s}$. Constraint~\eqref{eq_cp:school_population_bound} makes
  sure that the school populations do not exceed, or fall below, some
  thresholds specified by a parameter $\alpha$ (set to 0.15 based on
  the same parent survey as described above). Note that
  constraint~\eqref{eq_cp:school_population_bound} needs to be
  satisfied for all scenarios $i = 1,..,I$.

\item \textbf{Travel Time
  Increases}. Constraint~\eqref{eq_cp:travel_time_bound} ensures that
  driving times to assigned schools do not exceed some relative threshold
  $\tau$ (set to 0.5, inferred from a
  prior parent survey administered as a part
  of~\cite{gillani2023redrawing}). Driving times
  $\bar{t}_{b,s}$ were estimated from the centroid of each Census block
  to each elementary school. 

\item \textbf{Contiguity}. Constraint~\eqref{eq_cp:contiguity} ensures
  reasonable zoning shapes, where blocks that are contiguous with
  respect to their assigned schools remain contiguous in any new
  zoning (see the Supplementary Materials
  of~\cite{gillani2023redrawing} for additional details on how this is
  enforced, inspired by~\cite{mehrotra1998contiguity}).
\end{enumerate}

\noindent
Assuming that students in their present schools ($\bar{s}_n^{actual}$) do not want another
school, the model always has a feasible solution, since it suffices to
assign students to their school.

Constraint programming was used for its modeling flexibility, though other approaches are possible~\cite{gurnee2021fairmandering,smc, autrey2019}.

\begin{figure}[!t]
\begin{mini!}
    {}
    {
        \sum\limits_{i =  1}^I d^i 
        \label{eq_cp:obj}
    }
    {\label{formulation:cp_c}}
    {}
    \addConstraint
    {
       d^i =
    }
    {   
        \frac{1}{2}~\sum\limits_{s \in S} ~\bigg| \frac{g_s^i}{\bar{g}_{total}} - \frac{c_s^i - g_s^i}{N - \bar{g}_{total}} \bigg| 
        \label{eq_cp:dissimilarity_dynamic}
    }
    {}
    \addConstraint
    {~}
    {i = 1,...I \notag}
    {}
    \addConstraint
    {
        c_s^i =
    }
    {   
        \sum\limits_{n = 1}^N \sum\limits_{s' \in S} z_{\bar{b}_n, s'} \cdot \mathds{1}[A_n^i(s') = s] 
        \label{eq_cp:school_population_dynamic}
    }
    {}
    \addConstraint
    {~}
    { \forall  s \in \mathcal{SH}, \enspace i = 1,...I \notag}
    {}
    \addConstraint
    {
        g_s^i =
    }
    {   
        \sum\limits_{n = 1}^N \mathds{1}[\bar{e}_n = 0] \sum\limits_{s' \in S} z_{\bar{b}_n, s'} \cdot \mathds{1}[A_n^i(s') = s] 
        \label{eq_cp:school_ses_population_dynamic}
    }
    {}
    \addConstraint
    {}
    { \forall  s \in \mathcal{SH}, \enspace i = 1,...I \notag}
    {}
    \addConstraint
    {
        \bar{c}_s^i \cdot
    }
    {   
         (1 + \alpha) \geq  c_s^i \geq  (1 - \alpha) \cdot \bar{c}_s^i
        \label{eq_cp:school_population_bound}
    }
    {}
    \addConstraint
    {}
    { \forall  s \in \mathcal{SH}, \enspace i = 1,...I \notag}
    {}
    \addConstraint
    {
         1 =
    }
    {   
        \sum\limits_{s \in \mathcal{SH}} z_{b,s} \label{eq_cp:assign_1_school}
    }
    {
        \forall b \in \mathcal{B}
    }
    \addConstraint
    {
        \bar{t}_{b, \bar{s}_b} \cdot
    }
    {   
       (1 + \tau) \geq \sum\limits_{s \in \mathcal{SH}} z_{b,s} \cdot \bar{t}_{b, s}
            \label{eq_cp:travel_time_bound}
    }
    {
       \forall b \in \mathcal{B}
    }
    \addConstraint
    {
        ~
    }
    {   
        \textsc{Ensure Contiguity} (b) \label{eq_cp:contiguity}
    }
    {
        \forall b \in \mathcal{B}
    }
    \addConstraint
    {
         ~
    }
    {   
        a_{b, s} \in \{0, 1\} ~ \forall b \in \mathcal{B}, s \in \mathcal{SH} \label{eq_cp:binary}
    }
    {}
\end{mini!}
\caption{The Constraint Programming Model.}
\label{fig:cp}
\end{figure}

\subsection{Machine Learning for School Choice Modeling}

The contextual distributions $\Pr[ S_n | \mathbf{z}, \mathbf{x}_n]$
are approximated by supervised machine learning (e.g., boosted trees
or multinomial logit models). The learning task can be seen as a
multi-class classification problem, where each class represents a
particular school that a student might attend. The model is trained
from a dataset $\{(\mathbf{f}_k,\mathbf{d}_k), \mathbf{y}_k \}_{k}$
where, for each training instance $k$, $\mathbf{f}_k$ are static
features related to demographics and geography, $\mathbf{d}_k$ are
dynamic features related to school zoning (together they form the feature matrix), and $\mathbf{y}_k$ is the students' current school
choices (i.e., $\bar{s}_n^{actual}$ for student $n$ in the 22-23
school year). The Supplementary Materials include additional technical details on these features.

\section{Computational Results} 
\label{sect:results}

This section presents the computational results for RWC. It first
focuses on evaluating the predictive capabilities of
$\mathcal{C}^{ml}$ for student school choices. The experiments employ
two widely-adopted machine learning algorithms: multinomial logit and
XGBoost \cite{chen2016xgboost}. For comparative analysis, two baseline choice models are
introduced. The performance of these models is then assessed in the
multi-class classification task introduced earlier. Secondly, the
optimization results are presented, showing the benefit of using RWC
on school redistricting. Finally, this section analyzes the
performance stability of the SAA method.

The results are produced using Python 3.10,
the scikit-learn (1.5.1) and XGBoost (2.1.1) packages for machine learning,
and the ``CP-SAT" solver for constraint programming, with 8 threads for each run. All computations were performed using the first CPUs available on a Linux High Performance Computing cluster with Intel Xeon Gold, Platinum, and AMD Epyc CPUs. 

\begin{table*}[!ht]
\centering
\resizebox{0.8\textwidth}{!}{ 
\begin{tabular}{ l r r r  r r r  r r r}
\toprule
& & &  & \multicolumn{3}{c}{Macro Averaged} & \multicolumn{3}{c}{Weighted Averaged} \\ \cmidrule(lr){5-7} \cmidrule(lr){8-10}
Choice Model &  acc. & top 3 acc. & top 5. acc & precision & recall & f1-score & precision & recall & f1-score \\
\midrule
$\mathcal{C}^{f}$ (baseline 1) & 0.6533 & - & - & 0.6534 & 0.6522 & 0.6428 & 0.6651 & 0.6533 & 0.6498 \\
$\mathcal{C}^{rb}$ (baseline 2) & 0.4606 & 0.7013 & 0.7686 & 0.5124 & 0.4610 & 0.4710 & 0.5338 & 0.4606 & 0.4810 \\ 
$\mathcal{C}^{ml\text{-}l}$ (logit) & 0.6917 & 0.8516 & 0.9060 & 0.6859 & 0.6854 & 0.6827 & 0.6925 & 0.6917 & 0.6894 \\
$\mathcal{C}^{ml\text{-}x}$ (XGboost) & 0.7421 & 0.8748 & 0.9200 & 0.7360 & 0.7354 & 0.7345 & 0.7422 & 0.7421 & 0.7410 \\
\bottomrule
\end{tabular}
}
\caption{Evaluation of different choice models. For $\mathcal{C}^{ml\text{-}l}$ and $\mathcal{C}^{ml\text{-}x}$, their top 3 acc. and top 5 acc. are averaged across the 10-fold validation test sets. All other metrics are computed on the summed confusion matrices across all cross-validation fold.}
\label{tab:ml_metrics}
\end{table*}

\subsection{Baselines for Choice Modeling}

This section specifies two baselines for choice modeling. The first
baseline $\mathcal{C}^{f}$ simply assigns students to their zoned
school. Since approximately 2/3 of
elementary students already follow this policy in the dataset, it is a
coarse simplification to apply this to all students. It is also a
valuable baseline for the machine learning models, since they
might trivially ``learn'' to predict that students will select their
zoned schools and thus be correct approx. 2/3 of the time.

The second baseline $\mathcal{C}^{rb}$ uses frequency information in
the dataset, i.e., how frequently students attend or opt-out of their
zoned schools. As mentioned before, students attend their zoned
school about 65\% of the time; a nearby magnet school 20\%
of the time; and another nearby school the rest of the time, since
families tend to prefer lower travel times to
school~\cite{gillani2023redrawing}. The choice model is given
by
\begin{equation}
\label{eq:dist_rb}
\resizebox{0.85\columnwidth}{!}{
    $\Pr[S_n = s \mid \mathbf{x}_n] = 
    \begin{cases}
        0.65 &  \textit{if} \quad s = s_n^{zone}\\
        \cfrac{0.2}{|\mathcal{SH}_{mgt} \cap \mathcal{SH}_{near, 12}|}  &  \textit{if} \quad s \in \mathcal{SH}_{mgt} \cap \mathcal{SH}_{near, 12} \\
        0.03   & \textit{if} \quad s \in \mathcal{SH}_{near, 5}
    \end{cases}
    $
}
\end{equation}
where $\mathcal{SH}_{mgt}$ represents the magnet schools, and
$\mathcal{SH}_{near,r}$ denotes the $r$ closest schools for student
$n$.\footnote{The number 12 was determined as the minimum number of
  nearby schools needed to be included to ensure, for each student,
  that at least one of them is a magnet school.} Note that $s_n^{zone}$ is a variable, which is critical to construct $\mathbf{x}_n$.

\subsection{Choice Modeling Results}

This section reports the machine learning results and their comparison
with the baselines $\mathcal{C}^{f}$ and $\mathcal{C}^{rb}$. The
machine learning models are denoted by $\mathcal{C}^{ml\text{-}l}$ and
$\mathcal{C}^{ml\text{-}x}$ for the multinomial logit and XGboost,
respectively. Both models are evaluated using 10-fold
cross-validation, randomly holding out 10\% of students each time as
the test set. For $\mathcal{C}^{ml\text{-}l}$, scikit-learn defaults
are used for all parameters except for the max iteration amount, which
is set to 2,000. $\mathcal{C}^{ml\text{-}x}$ is trained with the
``multi'' objective and ``mlogloss'' metric; the model implements
early stopping with a validation set (15\% of the training set) to
help prevent over-fitting; and the learning rate and the maximum depth
of a tree are tuned to 0.1 and 6 after hyper-parameters tuning with
$[0.08, 0.1, 0.12]$ and $[4, 6, 8, 10, 12]$ on the validation set,
respectively.

The model outputs $\mathbf{y}^{pred}$ are compared to the ground
truth $\mathbf{y}$ for test-set students to evaluate
accuracy. Table~\ref{tab:ml_metrics} shows accuracy metrics for the
different choice models. In general, $\mathcal{C}^{ml\text{-}x}$
outperforms all the other models. Still, the accuracy and macro-averaged
F1 scores of 0.74 and 0.73, respectively, suggest that school choice
prediction is a challenging task. One reason for this may be the large
classification space of 41 schools. Another is that many unobservable
factors might influence a family's decision to select any particular
school. This is confirmed by prior research in a district where
families submit ranked preferences for schools: the study found
that multinomial logit models were correct less than 50\% of the time in
predicting a student's top-choice school~\cite{pathak2017demand}. The
current paper offers a large improvement over this previous finding --
improvements that may be attributed to the quality of the dataset and
feature engineering. Details on the specific features included in the
models are included in the Supplementary Materials. Exploring other
features is also a valuable direction for future research.

Interestingly, the top-3 and top-5 accuracy results are much higher
for $\mathcal{C}^{ml\text{-}x}$: 87\% of the time, the student's
actual school selection is captured by the top-3 highest-probability
schools indicated by the model. For top-5, this value is at
92\%. Inspecting the test set probabilities assigned to the top-3
schools across the 10 cross-validation folds shows that, on average,
the highest-probability school receives a probability mass of 76\%,
followed by 10\% for the second highest-probability school, followed
by 4\% for the third (with standard errors close to zero).

\subsection{Redistricting with choice}

The CSO of RWC was evaluated for the following three choice models:
$C^{f}$, $C^{rb}$, and $C^{ml\text{-}x}$. These are denoted by R for
Redistricting), FR (for Frequency-based Redistricting), and RWC
respectively. $C^{ml\text{-}l}$ was not considered due to its lower
performance in comparison $C^{ml\text{-}x}$. Before the redistricting steps, $C^{ml\text{-}x}$ is retrained
using the same parameters as the best-fitting model described earlier,
but with an 85\%-15\% split for the training and validation sets (no
test set is required since the model's primary use-case now is
inference). For the SAA method, the number of scenarios $I$ is set to
30 and the CP solver is configured to terminate after a 12-hour
solving period (none were able to prove optimality).

\begin{table*}[!t]
\centering
\resizebox{\textwidth}{!}{ 
\begin{tabular}{ l r r r r r r r}
\toprule
& & & \multicolumn{3}{c}{\textbf{\# Rezoned (\% Rezoned)}} & & \\ \cmidrule(lr){4-6}
\textbf{Method} & \textbf{\makecell[c]{\# SAA \\ Scenarios \\ (I)}}& \textbf{\makecell[r]{Dissimilarity \\ (standard error)}} & \makecell[r]{\textbf{Lower-SES} 
\\ \textbf{students}} & \makecell[r]{\textbf{All students}} & \makecell[r]{\textbf{Census Blocks}}& \makecell[r]{\textbf{Average} \\\textbf{\# Students opting out} \\ \textbf{of zoned school}} & \makecell{\textbf{Average} \\ \textbf{driving} \\ \textbf{time (min)}} \\
\midrule
Current & -& 0.596 (-) & - & - & - & 7,731 (34.67\%) & 7.54 \\
R    & - & 0.403 (-) & 1,870 (20.69\%) & 5,369 (24.07\%) & 1,902 (29.84\%) &  5,731 (25.70\%)& 7.44\\
FR &30 & 0.462 ($4.0\times10^{-4}$) & 1,870 (20.69\%) & 4,397 (19.72\%) & 1,512 (23.73\%) & 7,433 (33.33\%) & 7.53 \\
RWC  & 30 &0.459 ($4.6\times10^{-4}$) & 3,465 (38.33\%) & 7,864 (35.26\%) & 2,307 (36.20\%) & 9,227 (41.37\%) & 7.88 \\
\bottomrule
\end{tabular}
}
\caption{Results from different redistricting methods. For FR and RWC,
  dissimilarity values are averaged over scenarios (with resulting
  dissimilarity index standard errors in parenthesis). Note that R is
  deterministic. Interestingly, both R and FR resulted in the rezoning
  of 1,870 lower-SES students. This similariy is coincidental as shown
  in the maps in the Supplementary Materials, which are quite different.}
\label{tab:rezone_statistics}
\end{table*}

Table~\ref{tab:rezone_statistics} shows the redistricting results from
these models. Note that none of the CP models were solved to
optimality within 12 hours due to the complexity of the problem. The
table also includes statistics for the current (status quo)
assignment. Model R represents the ``best-possible'' redistricting by
ignoring that students may opt-out from their boundary-assigned
schools. It decreases dissimilarity by 32\% (from 0.596 to 0.403)
across the district. RWC, which integrates the machine learning choice
model and is grounded in a much more realistic setting, produces a
comparably impressive reduction in SES segregation of 23\%. This
reduction is nearly 2x larger than reductions discovered in previous
school redistricting work focused on fostering racial/ethnic
integration, but without choice modeling~\cite{gillani2023redrawing}.
Interestingly, RWC also outperforms FR, which uses a much less
realistic choice model. Travel times across all models, as well as the
status quo, are similar---suggesting that students opting out of their
zoned schools still generally select ones nearby.

Further inspecting the results in Table~\ref{tab:rezone_statistics}
suggests that the models appear to be making different decisions about
which students would opt out of their zoned schools. As expected, by
virtue of how the rule-based choice model is defined for FR,
approximately 1/3 of students are estimated to opt-out of their zoned
school even after redistricting (which closely mirrors ground-truth
opt-out rates). In the RWC model, more than 40\% are estimated to
opt out---however, as shown in Supplementary Materials, opt-out rates
under this model seem to be consistent across
demographics. Interestingly, as the columns labeled ``\# Rezoned''
show, RWC also tends to rezone more Census Blocks, lower-SES students,
and students overall compared to the other models. These trends may
reflect the need to make more drastic boundary changes to account for
more realistic patterns of choice that might re-segregate schools, as
choice sometimes
does~\cite{whitehurt2017segregation,candipan2019neighborhoods}. Maps
with results from the different redistricting models can be found
in the Supplementary Materials. The findings show that RWC
is not trivially replicating R (predicting students will attend
their zoned school).

The stability of the SAA method for RWC is evaluated through
additional experiments. Specifically, 100 RWC runs are conducted with
$I$ = 30, resulting in an average outcome of 0.466 dissimilarity score
and a standard error of 0.0012. Further investigation involved
increasing $I$ to 50, with 100 RWC runs performed under these
conditions. The results showed a higher average dissimilarity of 0.495 and a
standard error of 0.0049---i.e., on average, the results are less effective in reducing segregation than when $I$ = 30. This may be due to the fact that while 61 out of the 100 runs produce
results below 0.47, which aligns with findings when $I$ =
30, the remaining runs result in much higher dissimilarity values, closer to the observed ground truth of 0.6. This behavior suggests that the observed variations are likely due
to challenges associated with optimizing the CSO model as $I$
increases, given that a larger value of $I$ also translates into a more complex optimization problem. Future research may develop more efficient
methods for solving the RWC model.

\section{Discussion and Conclusion} \label{sect:discussion}

This paper presents RWC, a novel joint redistricting and choice
modeling framework that uses {\em a contextual stochastic constraint
  program} to minimize district-wide SES segregation. RWC derives the
school choice model using a machine learning model, leveraging
features of students, Census Blocks, and district schools. With a
realistic choice model, RWC results in large reductions
(23\%) in SES segregation, suggesting it may help foster more
integrated schools even after accounting for families who alter
their school selections in response to such changes. RWC can thus
support both researchers and practitioners in realizing the renewed promise of fostering diverse schools~\cite{jacobson2023fostering}.

Still, the paper has a number of limitations, which may serve as
important directions for future work. Perhaps the most obvious is the
choice model's performance. Anticipating school choice has been a
challenging task for other researchers
like~\cite{pathak2017demand}. Nevertheless, an exciting opportunity
for future work includes identifying more accurate prediction
approaches -- for example, through synthetic or historical data
augmentation; a richer feature space that additionally factors in
qualitative preferences (e.g. parents' online school
reviews~\cite{gillani2021reviews}); and more sophisticated machine
learning models.

Another limitation is that this study only models within-district
choice. Families may leave district schools for private, public
charter, or other options following a student assignment policy
change~\cite{macartney2018boards,reber2005flight,nielsen2020denmark,mervosh2021minneapolis}. This
out-of-district choice is more difficult to model due to limited
ground truth data. Still, finding ways to anticipate this choice may
yield a more accurate picture of a given redistricting's expected
impact on integration.

Even the present within-district redistricting-with-choice approach
might be improved by iterating the choice and redistricting models. At
present, this study does not simulate a true ``game'' between boundary
changes / impacts on school populations and the choices that families
might make. The study estimates choices as a function of changes to a
student's zoned school assignment; but in practice, as prior work on
neighborhood ``tipping''~\cite{card2008tipping} has suggested, a
family's choice may also depend on how aspects of a school that can
only be measured post-rezoning (like their demographics) might
change. Modeling this more complex interplay would require more
computationally demanding bi-level optimization approaches.

Finally, future work may explore how the RWC model makes
decisions, and why it tends to produce more drastic boundary
changes. While the model achieves larger reductions in segregation, it
leads to more students being rezoned and more opt-outs. This reflects
prior literature like~\cite{nielsen2020denmark} suggesting that
families may opt-out of new school assignments if they are
unfavorable. Understanding the predictors of opt-outs through
interpretable machine learning methods like~\cite{lundberg2020shap}
may help districts better understand and prepare for community
responses to integration policies.

Fostering integrated schools is a perennial challenge across US public
school districts with large implications for access to quality
educational environments and future life outcomes. Anticipating the
potential impacts of integration policies requires factoring in both
top-down policy priorities and bottom-up, emergent responses to
potential policy changes. RWC offers one starting point for jointly
modeling these forces to help advance the development of more
equitable student assignment policies in the coming years.

\section*{Acknowledgements}
This work was partially supported by Winston-Salem/Forsyth County Schools via the US Department of Education Fostering Diverse Schools Grant, the Overdeck Family Foundation, and NSF
AI Institute for Advances in Optimization Grant 2112533.
\bibliography{aaai25}

\appendix

\clearpage
\onecolumn
\section{Current Attendance of Each School}
\begin{figure*}[!ht]
    \centering
    \includegraphics[width=\textwidth]{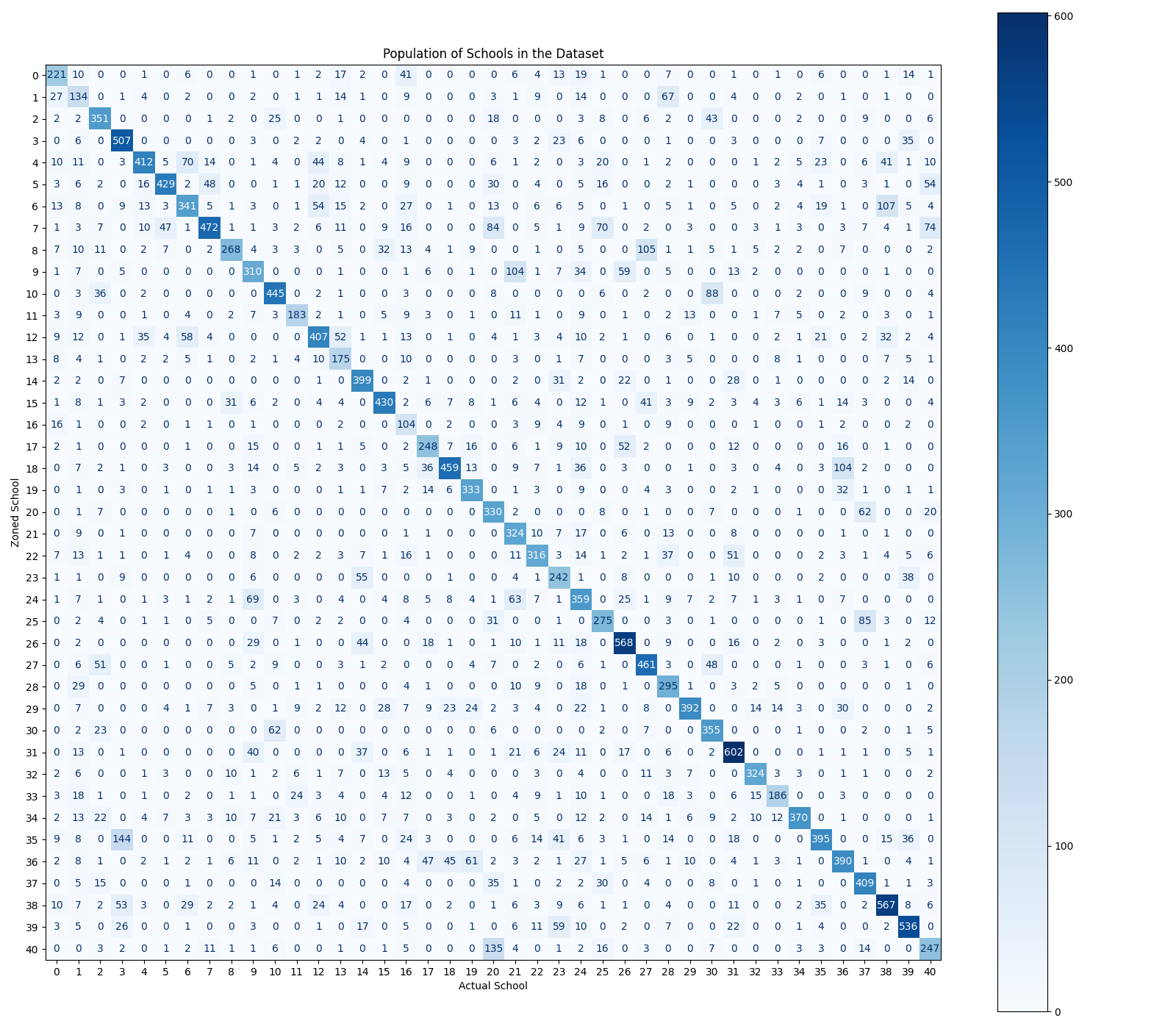}
    \caption{Matrix illustrating each schools' attendance during the 2022-2023 school year. Rows indicate the zoned school; columns indicate the actually attended schools. The diagonal shows a high percentage because most students attend their zoned schools. Off-diagonal entries are generally sparse, which likely makes the machine learning problem more challenging.}
    \label{fig:data_popluation}
\end{figure*}

Based on the dataset introduced in the main manuscript (see Section~\ref{sect:problem_method}), the matrix in Figure~\ref{fig:data_popluation} illustrates the population distribution across the 41 schools in this study, taking into account of students' zoned and actual schools.

\clearpage
\section{Features for Machine Learning and Contextual Information}
Table~\ref{tab:features} lists all of the features and their data types used in the machine learning-based choice model $\mathcal{C}^{ml}$. The school year for the classification problem is 2022-2023; that is to say, the actual schools of the students of that school year will form the label vector $\mathbf{y}$. In Table~\ref{tab:features}, the values of the bottom part features (below the line) will be changed once $s_n^{zone}$ is changed to another value, this mechanism is significant applying the Contextual Stochastic Optimization (CSO) framework. Additionally, Table~\ref{tab:features} utilizes the ``choice zone'' (denoted as $\mathcal{CZ}$) concept, defined by the school district. Each choice zone includes multiple schools, allowing students to attend any school within the same choice zone as their zoned school while retaining all the benefits, such as school bus services. It is important to note that each school can belong to multiple choice zones. Lastly, the school ratings data used in this study are extracted from the school rating website GreatSchools.org. 
\begin{table*}[!ht]
    \centering
    \begin{tabularx}{\textwidth}{X l X}
        \toprule
        \textbf{Feature Name} & \textbf{Feature Type} & \textbf{Additional Details} \\ \midrule
        race & Categorical & race $\in$ \{black, white, asian, native, hispanic, pacific-islanders, multiple \} \\
        \# students in census block $b_n$ & Discrete & - \\ 
        \% students in census block $b_n$ & Continuous & Compared to the total number of students $N$ \\ 
        \# students of each race in census block $b_n$ & Discrete  & - \\ 
        \% students of each race in census block $b_n$ & Continuous  & Compared to the total number of students of this race in the dataset \\ 
        travel time to school $s$ & Continuous & - \\
        travel distance to school $s$ & Continuous & - \\
        SES Level & Ordinal & This is based on the students' residential areas, not on households \\
        grade level & Ordinal & Grade level at 22-23 school year \\
        school $s$ is a magnet school & Binary & $\forall s 
        \in \mathcal{SH}$ \\      
        new student to the school system & Binary & If the student has no record before the 2022-2023 school year. \\
        has siblings in elementary school & Binary & consider 22-23 school year \\
        previously went to the same school with their siblings & Binary & Using data before 22-23 school year \\
        has opted-out in previous school years & Binary & Using data before 22-23 school year \\
        has opted-out to a magnet school in previous school years & Binary & Using data before 22-23 schoo l year \\
        went to multiple schools in previous years & Binary & Using data before 22-23 school year \\

        \midrule
        zoned school $s_n^{zone}$ & Categorical & - \\
        zoned school belongs to choice zone $o$ & Binary & $\forall o \in \mathcal{CZ}$ because each school can belong to more than one choice zone\\ \\
        zoned school $s_n^{zone}$ and school $s$ are in the same choice zone & Binary & $\forall s \in \mathcal{SH}$\\
        different types of rating of school $s$ compared to the zoned school $s_n^{zone}$ & Continuous & $\forall s \in \mathcal{SH}$, the rating of school $s$ divided by the rating of school $s_n^{zone}$. The type of GreatSchools.org ratings considered are \{overall, test, progress, and equity\} \\
        \bottomrule
    \end{tabularx}
    \caption{The list of feature used to create $\mathbf{x}_n$ for student $n$. The static features above the mid horizontal line belong to $\mathbf{f}_n$, while those features below belong to $\mathbf{d}_n$ (based on variable $s_n^{zone}$).}
    \label{tab:features}
\end{table*}

\clearpage

\section{Analysis on District Maps after Rezoning}

This section presents some additional results on the rezoned school districts. Note that maps shown in this sections are hypothetical and not under consideration for the Fostering Diverse Schools project.

\begin{figure*}[!ht]
    \centering
    \begin{subfigure}[b]{0.40\textwidth}
        \includegraphics[width=\textwidth]{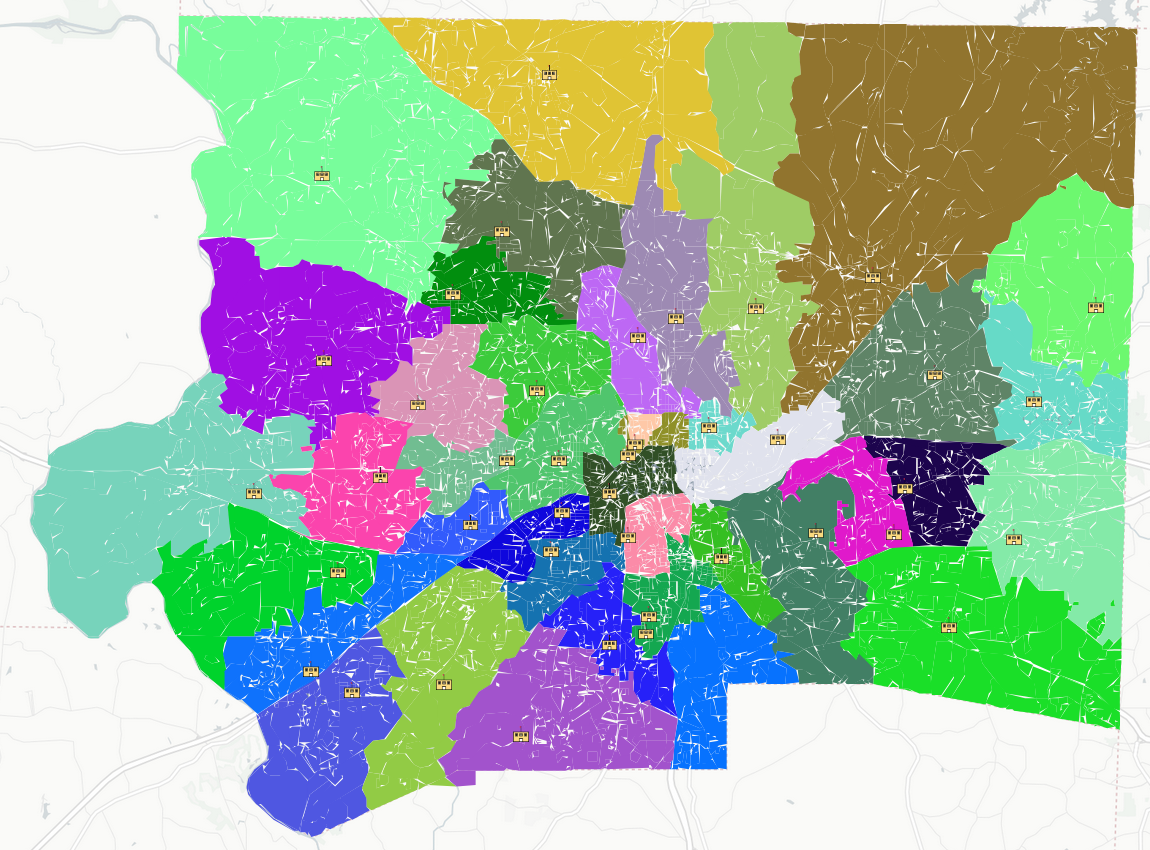}
        \caption{Current}
        \label{subfig:map_current}
    \end{subfigure}
    \begin{subfigure}[b]{0.40\textwidth}
        \includegraphics[width=\textwidth]{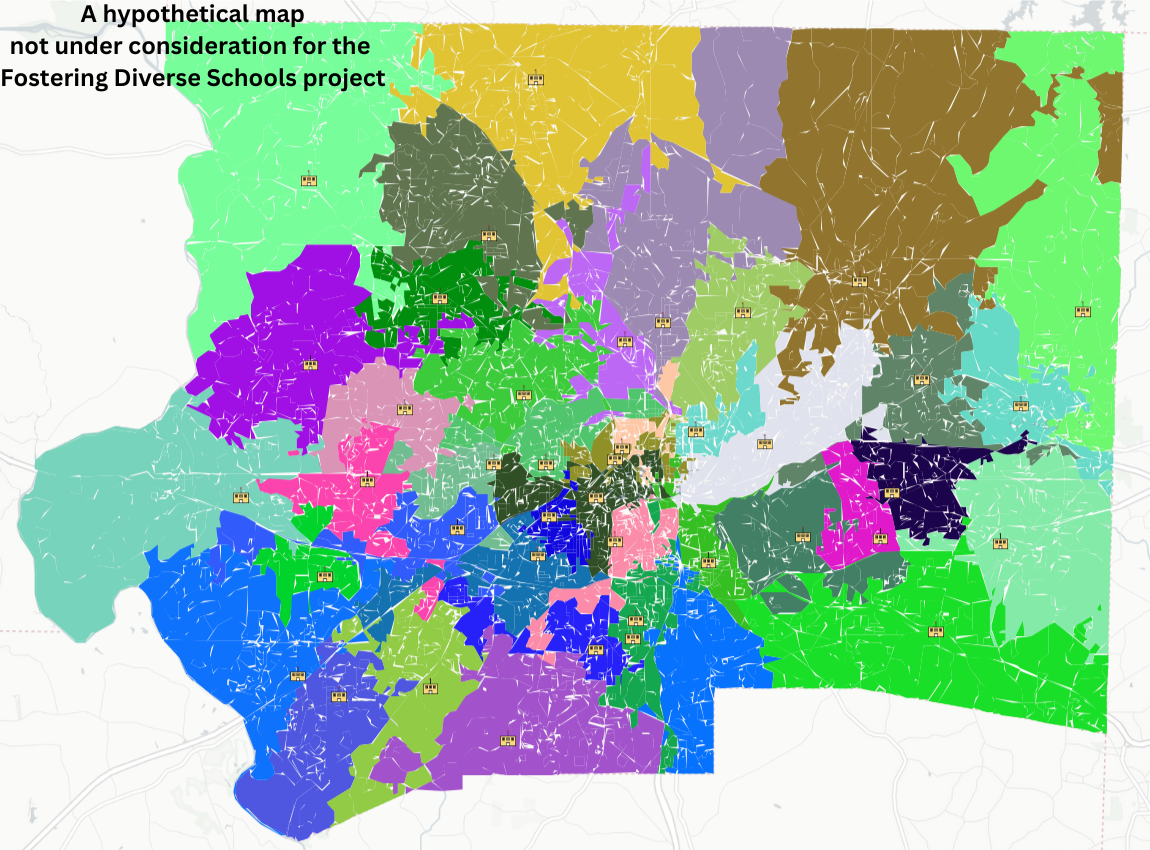}
        \caption{R}
        \label{subfig:map_baseline}
    \end{subfigure}
    \begin{subfigure}[b]{0.40\textwidth}
        \includegraphics[width=\textwidth]{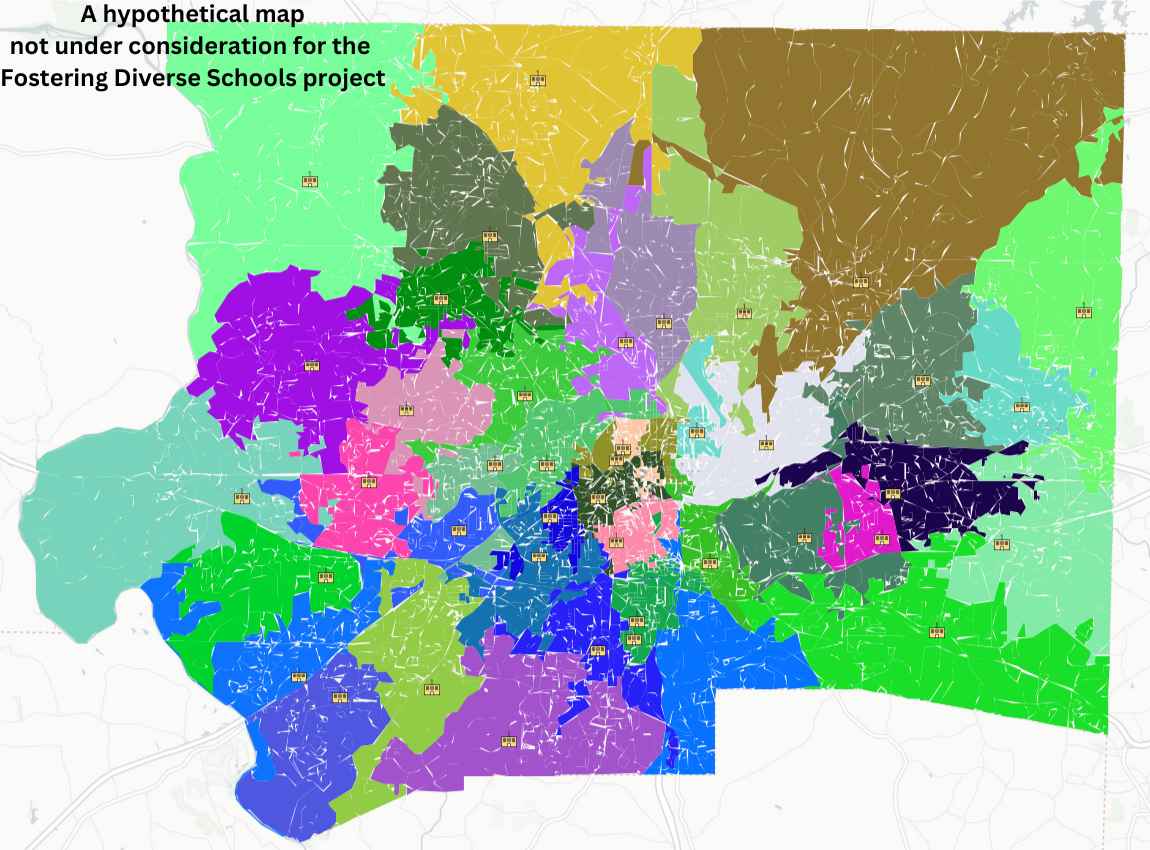}
        \caption{FR}
        \label{subfig:map_rb_cp}
    \end{subfigure}
    \begin{subfigure}[b]{0.40\textwidth}
        \includegraphics[width=\textwidth]{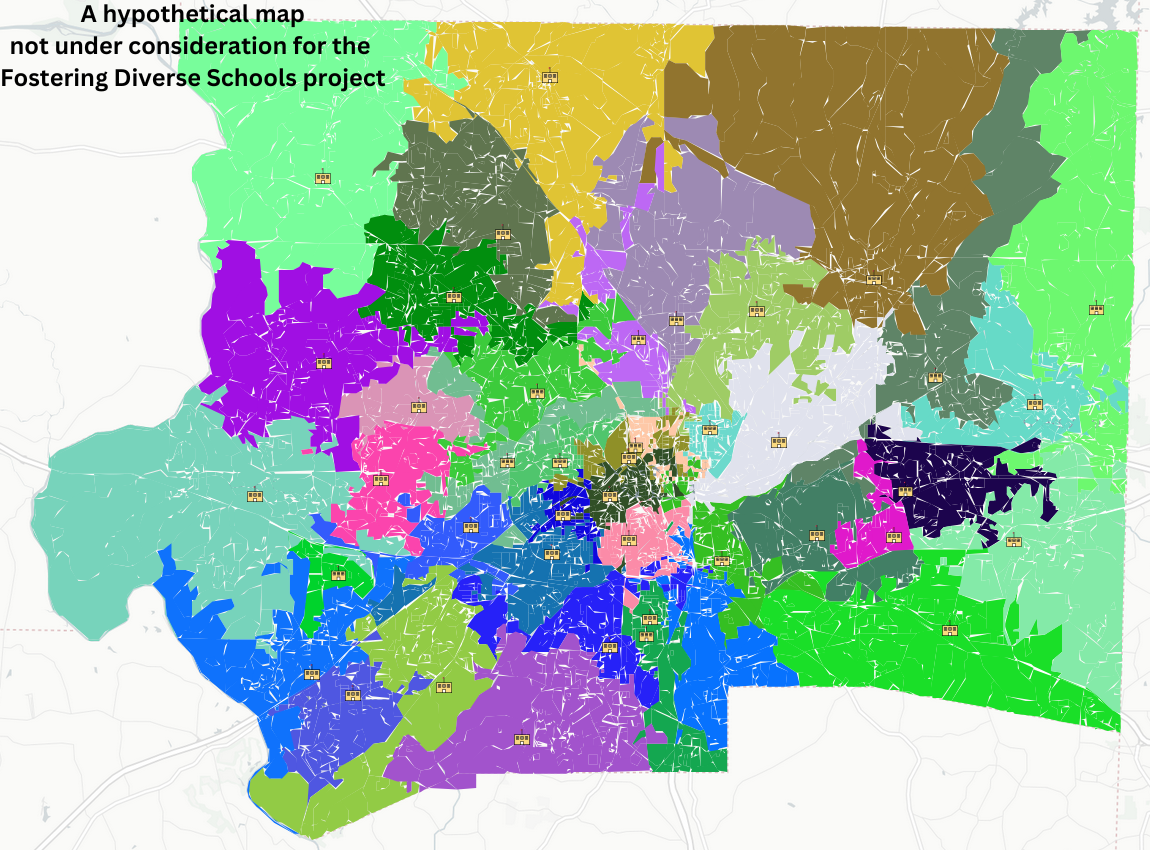}
        \caption{RWC}
        \label{subfig:map_rwc_cp}
    \end{subfigure}
\caption{\textbf{Maps shown here are hypothetical and not under consideration for the Fostering Diverse Schools project.} Elementary school attendance boundaries in the district's status-quo (a) and after applying each of the methods in the main document. Colors represent the same schools across maps.}
\label{fig:map_rezone}
\end{figure*}

\paragraph{Redistricted Attendance Boundaries}
Figure~\ref{fig:map_rezone} illustrate the attendance boundaries of each school under different methods. Note that some regions may visually appear to be discontinuous, but may be connected through small adjacent geographies (like rivers). The supplementary materials of~\cite{gillani2023redrawing} offer additional insights into how the contiguity constraints might visually manifest in resulting maps. 

\begin{figure*}[!ht]
    \centering
    \begin{subfigure}[b]{0.45\textwidth}
        \includegraphics[width=\textwidth]{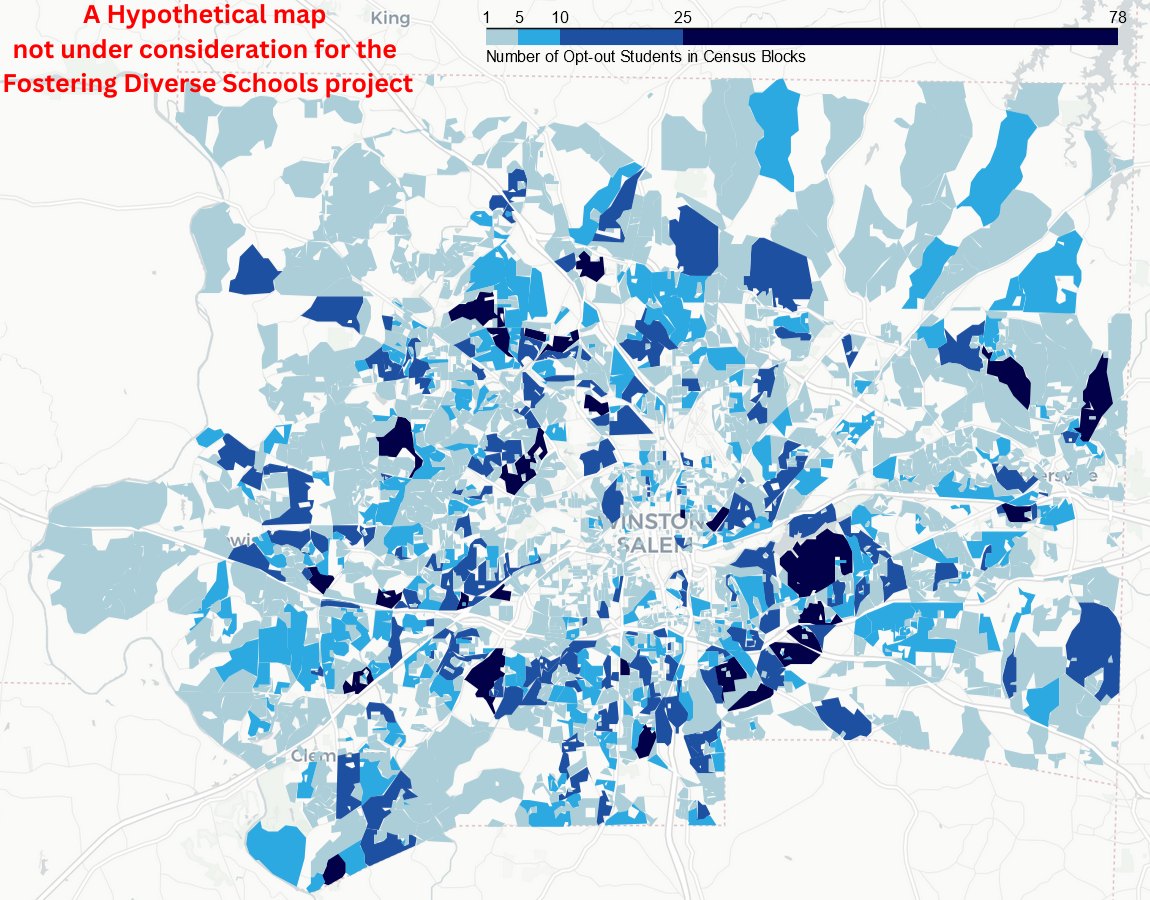}
        \caption{\# opt-out students}
        \label{subfig:map_rwc_opt_out_count}
    \end{subfigure}
    \begin{subfigure}[b]{0.45\textwidth}
        \includegraphics[width=\textwidth]{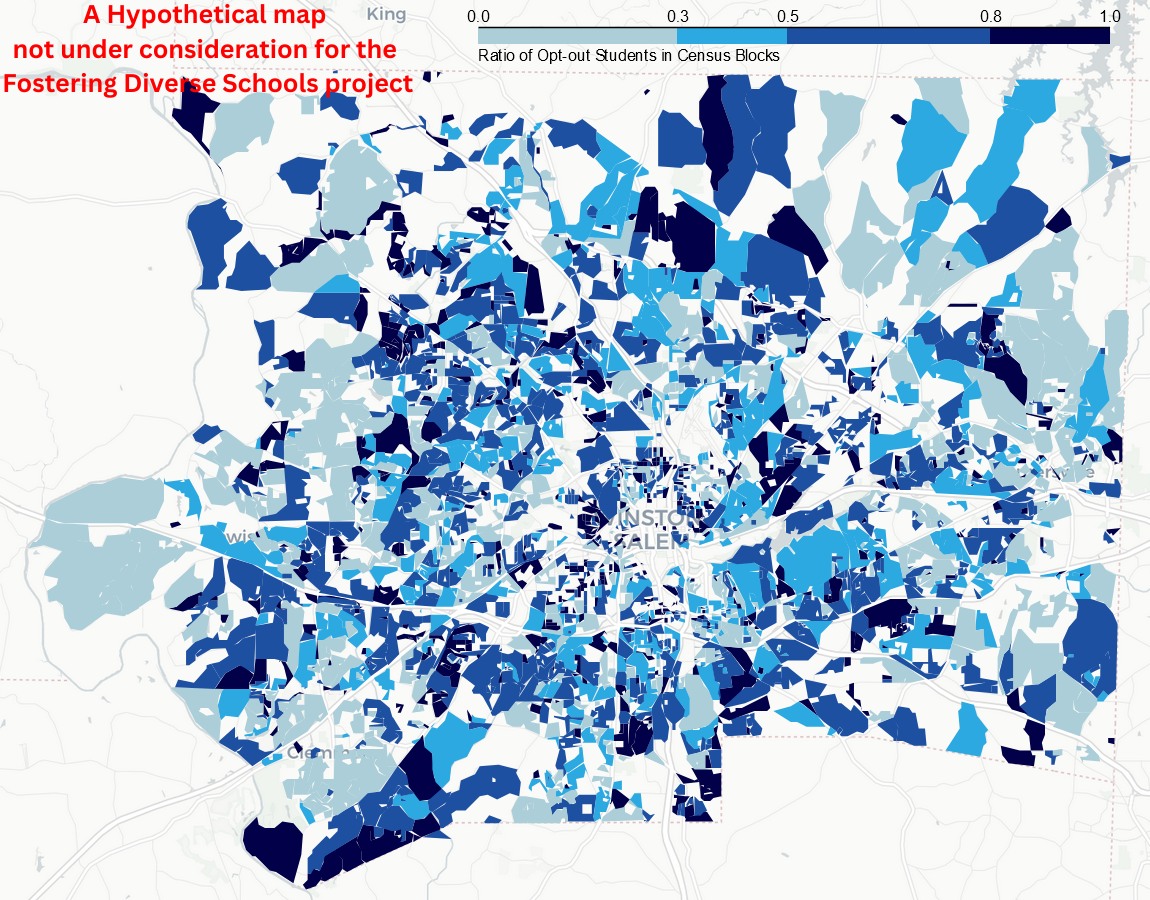}
        \caption{opt-out students ratio}
        \label{subfig:map_rwc_opt_out_ratio}
    \end{subfigure}
\caption{\textbf{Maps shown here are hypothetical and not under consideration for the Fostering Diverse Schools project.} The maps present the number and ratio of opt-out students in each residential census block, with values averaged over the 30 ($I$) counterfactual instances. Darker colors indicate a higher count or ratio of opt-out students within the census blocks. Census blocks without colors either have no elementary students (such as rivers) or no students opting out.}
\label{fig:map_opt_out}
\end{figure*}

\paragraph{Opting-out Students}
The locations of opt-out students are shown in Figure~\ref{fig:map_opt_out}. For each census block, the number of opt-out students is averaged over 30 instances, and then the opt-out ratio for each block is computed. To gain a better understanding of the demographics of the opting-out students, those who opt-out their school assignment 15 or more times across 30 instances were analyzed, totaling 5,191 students. Among these 5,191 students, 34.81\% (1,807) are white, and 40.28\% (2,091) are from lower-SES residential areas. These figures are comparable to the overall dataset, where 32.56\% (7,261) of students are white, and 42.02\% (9,376) are from lower-SES backgrounds.

\begin{figure*}[!ht]
    \centering
    \begin{subfigure}[b]{0.45\textwidth}
        \includegraphics[width=\textwidth]{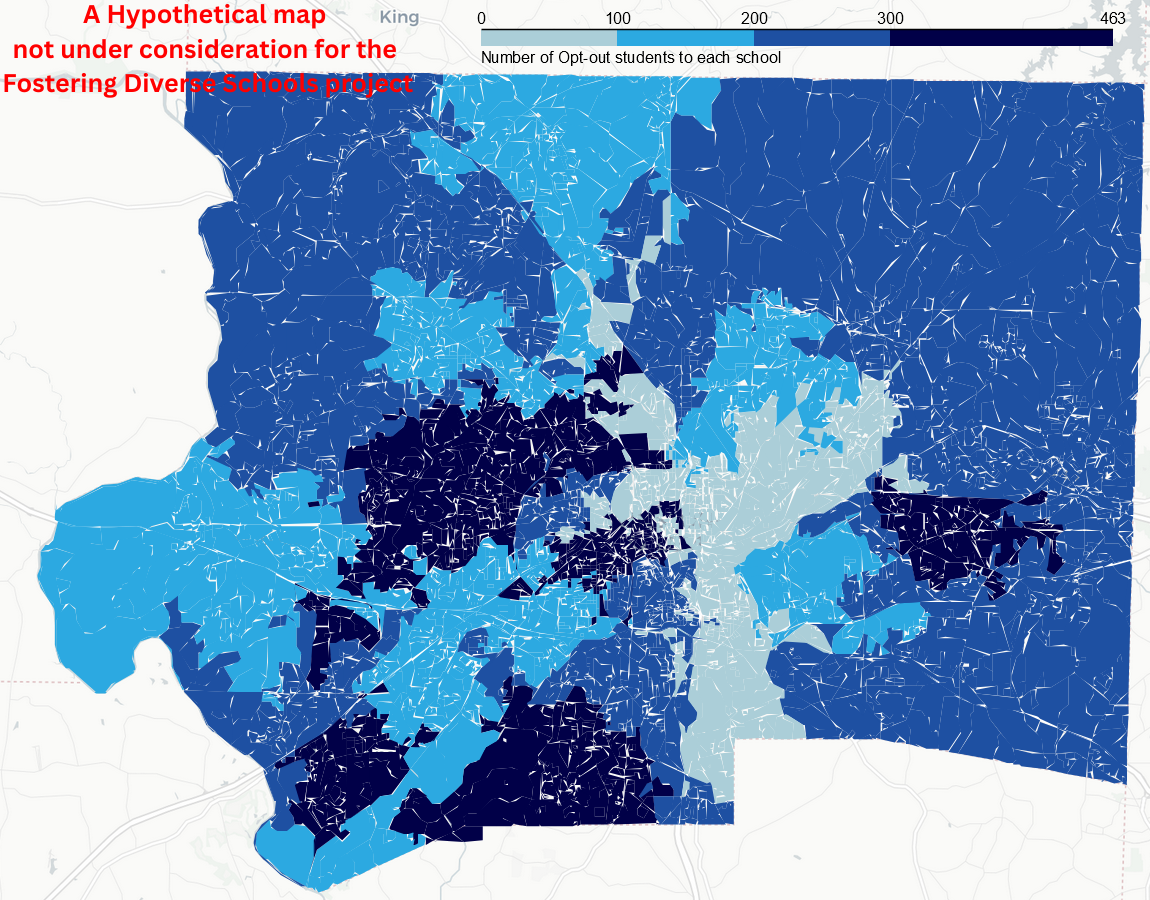}
        \caption{\# opt-in students to each school}
        \label{subfig:map_rwc_opt_out_to_schools}
    \end{subfigure}
    \begin{subfigure}[b]{0.45\textwidth}
        \includegraphics[width=\textwidth]{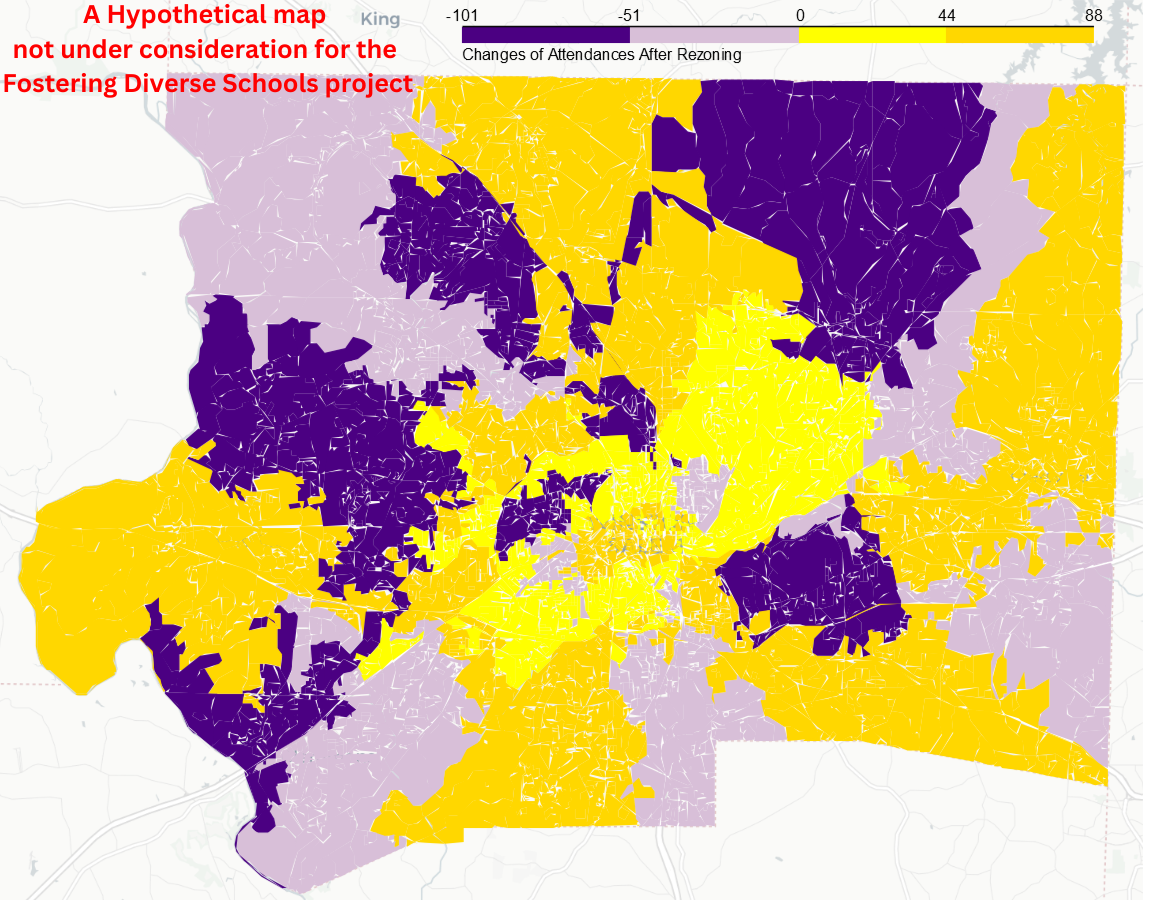}
        \caption{\# student changes after rezoning}
        \label{subfig:map_rwc_school_attendance_changes}
    \end{subfigure}
\caption{\textbf{Maps shown here are hypothetical and not under consideration for the Fostering Diverse Schools project.} The maps display the number of opt-in students to each school and the changes in student numbers after rezoning, averaged over the 30 instances. In Figure~\ref{subfig:map_rwc_opt_out_to_schools}, darker colors indicate schools that attract more opting-in students. In Figure~\ref{subfig:map_rwc_school_attendance_changes}, purple and yellow colors indicate schools that lose and gain students after rezoning, respectively.
}
\label{fig:map_school_changes}
\end{figure*}

\paragraph{Results on Schools}
Figure~\ref{fig:map_school_changes} presents two maps illustrating which schools attracted more opt-out students and how the student population at each school changed after rezoning with the RWC model. Among the top five schools with the highest number of opt-in students, two are magnet schools (ranked 2 and 4). Interestingly, the schools ranked 3 and 5, despite having students opt-in, still experienced a small net loss in student population compared to before rezoning.
\end{document}